\documentclass[a4paper,11pt]{article}
\usepackage{pos}

\title{Angular momentum distribution for a quark dressed with a gluon: different decompositions }

\author[a]{Ravi Singh}
\author[a]{Sudeep Saha}
\author*[a]{Asmita Mukherjee}
\author[b]{Nilmani Mathur}

\affiliation[a]{Department of Physics, Indian Institute of Technology Bombay\\
  Powai, Mumbai 400076, India}
\affiliation[b]{Department of Theoretical Physics,
 Tata Institute of Fundamental Research \\ 
 Colaba, Mumbai 400005, India}

\emailAdd{30004559@iitb.ac.in}
\emailAdd{sudeepsaha@iitb.ac.in}
\emailAdd{asmita@phy.iitb.ac.in}
\emailAdd{nilmani@theory.tifr.res.in}

\abstract{We present a recent calculation \cite{Singh:2023hgu} of the quark and gluon contributions to the angular momentum of a composite spin -$1/2$ state in QCD. The state we consider is a quark dressed with a gluon, and we use the two-component framework in light-front Hamiltonian QCD. We compare the results from different decompositions available in the literature. We also present the angular momentum distributions.}

\FullConference{%
  31st International Workshop on Deep Inelastic Scattering (DIS 2024)\\
  8 - 12 April 2024\\
  Grenoble, France
}

\begin{document}
\maketitle

\section{Introduction}
Experimental findings show that only a fraction of a nucleon's spin comes from the spin of its quarks and gluons \cite{EuropeanMuon:1989yki, deFlorian:2014yva}. The remaining portion is attributed to the orbital angular momentum (OAM) of these constituents, due to the nucleon's relativistic nature \cite{Dudek:2012vr}. However, the theoretical understanding of how nucleon spin arises from the sum of these angular momenta remains ambiguous \cite{Leader:2013jra}. Historically, one challenge was the gauge-invariant separation of gluon angular momentum into intrinsic and orbital components. Recent polarized scattering experiments have provided insights into the intrinsic spin of gluons within nucleons, necessitating gauge-invariant observables to interpret these findings effectively. It has been theoretically established that gluon angular momentum can be further divided gauge-invariantly into spin and orbital components, introducing a gauge-invariant "potential angular momentum" term \cite{Burkardt:2012sd} that can influence the definition of quark or gluon OAM. Thus many different decompositions of nucleon spin exist, each offering unique perspectives on angular momentum distribution. Although these decompositions agree on the total nucleon spin, they differ at the density level due to surface terms. Understanding these distinctions is essential for proper interpretation of the spatial distribution of nucleon angular momentum.

\section{Angular Momentum Density Decompositions}
Decompositions of angular momentum in QCD are classified into kinetic and canonical categories. The kinetic class includes the Belinfante, Ji, and Wakamatsu decompositions, with the latter being a gauge-invariant extension (GIE) of the Ji decomposition. The canonical class includes the Jaffe-Manohar decomposition and its GIE by Chen et al. This classification was enabled by Wakamatsu's covariant generalization of the QCD angular momentum tensor into five gauge-invariant terms: quark and gluon spin, orbital angular momentum (OAM), and potential angular momentum. Potential angular momentum can be added to either quark or gluon OAM, creating the canonical and kinetic families. While these decompositions yield the same total angular momentum when integrated, they differ at the density level due to superpotential (surface) terms that vanish upon integration. Thus, total angular momentum density cannot be simply interpreted as a sum of OAM and spin density. More details can be found in section 5.2.3 of Ref. \cite{Leader:2013jra}.

\section{Angular Momentum Distribution in Front Form}
We calculate intrinsic spin and OAM densities in the light-front coordinates. These distributions are analyzed in a 2D plane orthogonal to longitudinal motion to avoid relativistic corrections associated with 3D densities. Light-front coordinates provide an elegant framework where quantities evaluated in the transverse plane exhibit Galilean symmetry. The impact-parameter distribution or the 2D spatial distribution of orbital angular momentum is
{ \begin{align}
    \langle L^z \rangle(\boldsymbol{b}^{\perp})= -i\epsilon^{3jk}\int \frac{d^2\boldsymbol{\Delta}^{\perp}}{(2\pi)^2}e^{-i\boldsymbol{\Delta}^{\perp} \cdot \boldsymbol{b}^{\perp}} \left[\frac{\partial \langle T^{+k}\rangle_{\text{LF}} }{\partial \Delta_{\perp}^{j}}\right], \label{Lz expr}
\end{align} }
where { $\boldsymbol{b}^{\perp}$} is the impact parameter and { $\langle T^{\mu\nu}\rangle_{\text{LF}}=\frac{\langle p^{\prime},\boldsymbol{s}|T^{\mu\nu}(0)|p,\boldsymbol{s}\rangle}{2\sqrt{p^{\prime+}p^+}}$}. Here, $T^{\mu\nu}$ is the energy-momentum tensor.
Similarly, we can define the spin distribution in light-front as
{ \begin{align}
    \langle S^z \rangle(\boldsymbol{b}^{\perp})&= \frac{1}{2}\epsilon^{3jk}\int \frac{d^2\boldsymbol{\Delta}^{\perp}}{(2\pi)^2}e^{-i\boldsymbol{\Delta}^{\perp} \cdot \boldsymbol{b}^{\perp}} \langle S^{+jk}\rangle_{\text{LF}}, \label{Sz expr}
\end{align}}
and the Belinfante-improved total angular momentum distribution as
{ \begin{align}
     \langle J_{\text{Bel}}^z \rangle(\boldsymbol{b}^{\perp})
= -i\epsilon^{3jk}\int \frac{d^2\boldsymbol{\Delta}^{\perp}}{(2\pi)^2}e^{-i\boldsymbol{\Delta}^{\perp} \cdot \boldsymbol{b}^{\perp}} \left[\frac{\partial \langle T^{+k}_{\text{Bel}}\rangle_{\text{LF}} }{\partial \Delta_{\perp}^{j}}\right].\label{Jz expr}
\end{align}}
In order to ensure consistency between the Belinfante, Ji and Jaffe-Manohar decomposition in regards to the total quark AM density, it is important to include the correction term to the Belinfante's total angular momentum. At the distribution level, it is given as \cite{Lorce:2017wkb}
\begin{align}
    \langle M^z_{q} \rangle(\boldsymbol{b}^{\perp})&= \frac{1}{2}\epsilon^{3jk}\int \frac{d^2\boldsymbol{\Delta}^{\perp}}{(2\pi)^2}e^{-i\boldsymbol{\Delta}^{\perp} \cdot \boldsymbol{b}^{\perp}} \Delta^l_\perp \frac{\partial \langle S^{l+k}_{\text{q}}\rangle_{\text{LF}}}{\partial \Delta^j_\perp}.  \label{Mz expr}
\end{align}

For performing the Fourier transform of the aforementioned distributions, we utilized a Gaussian wave packet state with a fixed longitudinal momentum and finite width, confined within the transverse momentum space \cite{Singh:2023hgu}.
To calculate the matrix elements of local operators, we consider a relativistic spin-$\frac{1}{2}$ state of a quark dressed with a gluon at one loop in QCD \cite{Harindranath:1998pd}. 
\begin{align}
    \nonumber |p,\lambda\rangle =& \psi_1(p,\lambda)b_{\lambda}^{\dagger}(p)|0\rangle 
 +\sum_{\lambda_1, \lambda_2}\int \frac{dk_1^{+}d^{2}\boldsymbol{k}_1^{\perp}}{\sqrt{2(2\pi)^3k_1^+}}\int \frac{dk_2^{+}d^{2}\boldsymbol{k}_2^{\perp}}{\sqrt{2(2\pi)^3k_2^+}}\psi_2\left(p,\lambda|k_1,\lambda_1;k_2,\lambda_2\right)\sqrt{2(2\pi)^3P^+}\\&\delta^{3}\left(p-k_1-k_2\right)b_{\lambda_1}^{\dagger}(k_1)a_{\lambda_2}^{\dagger}(k_2)|0\rangle ,
\end{align}
where, {$\psi_1(p,\lambda)$} is normalization, {$\psi_2\left(p,\lambda|k_1,\lambda_1;k_2,\lambda_2\right)$} is the probability amplitude of finding a quark and gluon with momentum (helicity) {$k_1(\lambda_1)$ \& $k_2(\lambda_2)$} respectively. We have used two-component formalism developed in light front gauge $A^{+}=0$ \cite{Zhang:1993dd}. 

This work contains only the quark part of the EMT and an ongoing work contains the gluon part of the EMT \cite{Saha:2024hgu}. We found that the off-diagonal matrix element of the quark part of kinetic EMT is zero. So effectively the kinetic EMT coincides with the canonical EMT for the quark part of EMT. This signifes the vanishing of the potential angular momentum.

The longitudinal component of kinetic and canonical OAM distribution of quarks:
\begin{align}
            \langle L^z_{\text{kin, q}} \rangle (\boldsymbol{b}_\perp) =&\frac{g^2 C_f}{72 \pi^2} \int \frac{d^2\boldsymbol{\Delta}^{\perp}}{(2\pi)^2}e^{-i\boldsymbol{\Delta}_\perp \cdot \boldsymbol{b}_\perp }\bigg[-7+\frac{6}{\omega}\left(1+\frac{2m^2}{\Delta^2}\right)\text{log}\left(\frac{1+\omega}{-1+\omega}\right)-6\text{log}\left(\frac{\Lambda^2}{m^2}\right)\bigg], \label{OAM_kin}
        \end{align}
and kinetic and canonical spin distribution of quarks:
{ \begin{align}
   \langle S^z_{\text{kin, q}} \rangle (\boldsymbol{b}_\perp)\nonumber =& \int \frac{d^2\boldsymbol{\Delta}^\perp}{(2\pi)^2}e^{-i \boldsymbol{\Delta}_\perp \cdot \boldsymbol{b}_\perp}  \int dx \bigg[ \frac{1}{2}+\frac{g^2 C_f}{16 \pi^2(1-x)} \bigg\{2x- \omega (1+x^2) \log{\left( \frac{1+\omega}{-1+\omega} \right)}\nonumber \\
            & -\left(\frac{1-\omega^2}{\omega}\right) x \log{\left( \frac{1+\omega}{-1+\omega} \right)} \bigg\} \bigg],
    \label{Spin_kin}
\end{align}}
where { $\omega = \sqrt{1+\frac{4m^2}{\Delta^2}}$},  $x$ is the light-front momentum fraction of the quark and { $\Lambda$} is the ultraviolet cutoff on the transverse momentum \cite{Singh:2023hgu}.
\\ 
Belinfante-improved quark total angular momentum distribution:
        \begin{align}
            \langle J^z_{\text{Bel,q}} \rangle (\boldsymbol{b}_\perp) &= \int \frac{d^2\boldsymbol{\Delta}^{\perp}}{(2\pi)^2}e^{-i\boldsymbol{\Delta}_\perp \cdot \boldsymbol{b}_\perp} \int dx \bigg[\frac{1}{2}-\frac{g^2 C_f}{16 \pi^2} \bigg\{\frac{1+x^2}{1-x} \log{\bigg(\frac{\Lambda^2}{m^2 (1-x)^2}\bigg)} \nonumber -\frac{2x}{1-x}\bigg\} \bigg] \\&+\frac{g^2 C_f}{16\pi^2\left(1-x\right){\Delta}^4 \omega^3}\nonumber\bigg[\bigg(8m^4\left(1-2x\right)\left(1-x\left(1-x\right)\right) +6m^2\left(1-\left(2-x\right)x\left(1+2x\right)\right)\Delta^2 \nonumber \\&+\left(1-\left(2-x\right)x\left(1+2x\right)\right)\Delta^4\bigg)\log\bigg(\frac{1+\omega}{-1+\omega}\bigg) -\omega \Delta^2 \bigg(4m^2\left(1-\left(1-x\right)x\right)+\left(1+x^2\right)\Delta^2 \nonumber \\ &+\left(1-\left(2-x\right)x\left(1+2x\right)\right)\left(4m^2+\Delta^2\right) \log\bigg(\frac{\Lambda^2}{m^2\left(1-x\right)^2}\bigg)\bigg)\bigg],
        \end{align}
 Distribution of the superpotential term associated with quark 
\begin{align}
\langle M^z_\text{q} \rangle & (\boldsymbol{b}_\perp) = \frac{g^2 C_{f}}{16 \pi^2} \int \frac{d^2 \boldsymbol{\Delta}_\perp}{(2\pi)^2} e^{-i\boldsymbol{\Delta}_\perp \cdot \boldsymbol{b}_\perp} \int \frac{dx}{(1-x)} \frac{1}{\omega^3 \Delta^{4}} \times \nonumber \\
&\left[ \omega \Delta^{2} \left( (4m^2 + \Delta^{2})(1+x^2) - 4m^2 x \right)  - 2m^2 \left( (4m^2 + \Delta^{2})(1+x^2) - 4m^2 x - 2x \Delta^{2} \right) \right].
\end{align}
In the light front gauge, the physical part of the gauge potential is the same as the full gauge potential. Thus, densities of all the components of the gauge invariant decompositions coincide with the corresponding densities in canonical (JM) and kinetic (Ji) decompositions respectively. 
\section{Numerical analysis}
\begin{figure}[ht]
\begin{minipage}{0.49\linewidth}
\centering
\includegraphics[scale=0.60]{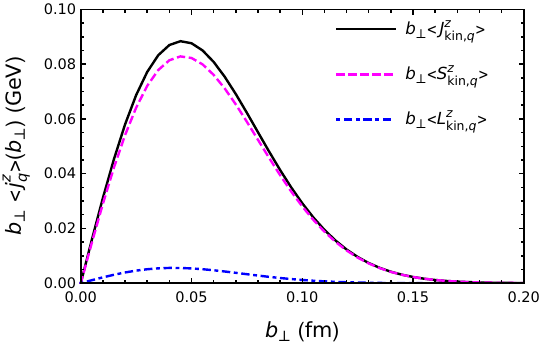}
\end{minipage}
\begin{minipage}{0.49\linewidth}
\centering
\includegraphics[scale=0.60]{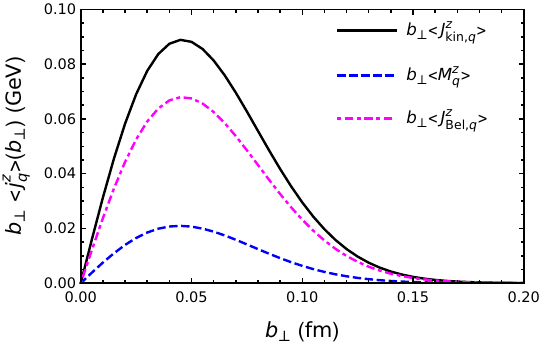}
\end{minipage}  
\caption{Longitudinal AM distribution of quarks as a function of impact parameter $|b_\perp|$. Left: Sum of $\langle S^z_{\text{kin,q}} \rangle$ and $\langle L^z_{\text{kin,q}} \rangle$ given by $\langle J^z_{\text{kin,q}} \rangle$. Right:$\langle J^z_{\text{kin,q}} \rangle$ is given by the sum of $\langle J^z_{\text{Bel,q}} \rangle$ and $\langle M^z_{\text{q}} \rangle$.\cite{Singh:2023hgu, Saha:2024hgu}.}
\label{qDist_graph}
\end{figure}

In this section, we analyze the longitudinal component of the angular momentum distribution of quarks and gluons. 
For the analysis, we have chosen: the quark mass $m = 0.3$ GeV, the coupling constant $g = 1$, the color factor $C_f = 1$, and $\Lambda = 2.63$ GeV. The y-axis is multiplied by a factor of $|\vec{b}_\perp |$ to correctly represent the data in radial coordinates.

In Fig. \ref{qDist_graph}, the plot on the left panel is a graphical representation of the longitudinal component of $\langle J^z_{\text{kin,q}} \rangle (b_\perp) = \langle L^z_{\text{kin,q}} \rangle (b_\perp) + \langle S^z_{\text{kin,q}} \rangle (b_\perp)$.
In the right panel, we show that the Belinfante total AM distribution, $\langle J^z_{\text{Bel,q}} \rangle (b_\perp)$ is not equal to the total AM density of the quark. 
The correction term $\langle M^z_{\text{q}} \rangle (b_\perp)$, ignored in the symmetric Belinfante EMT, must be added to $\langle J^z_{\text{q}} \rangle (b_\perp)$ to ensure the same total AM distribution in both decompositions

As we are using the QCD Hamiltonian, $J_{q}$ is not conserved, resulting in a scale or a cutoff dependence of the components \cite{Ji:2010zza}. Thus, by taking different values of the cutoff $\Lambda$, the expected equality $J_{\text{kin,q}}^{z}=J^{z}_{\text{Bel,q}}+M^{z}_q$  does not hold. 
So we have chosen a suitable value for the cut-off $\Lambda=2.63$ GeV so that this equality holds. The analysis of the cutoff dependence of distributions can be found in \cite{Singh:2023hgu}.

\section{Conclusions}
In this study, we analyzed the angular momentum distributions of a quark within a light-front dressed quark state using the Drell-Yan frame and two-component QCD in the light-front gauge $A^{+}=0$. Neglecting boundary terms or superpotentials led to discrepancies in total angular momentum density across different decompositions. We also found that the potential angular momentum term is zero, as expected, since no torque can exist between constituents in a two-body system like the dressed quark state.
\section{Acknowledgement} 
A. M. thanks the organisers of DIS2024 for the invitation. We acknowledge the funding from the Board of Research
in Nuclear Sciences (BRNS), Government of India, under
sanction No. 57/14/04/2021-BRNS/57082. A. M. thanks
SERB-POWER Fellowship (SPF/2021/000102) for financial support.

\bibliographystyle{JHEP}
\bibliography{references}


\end{document}